\documentstyle[prl,aps]{revtex}

\begin{document}
\author{Jian-Qi Shen $^{1,}$$^{2}$ \footnote{E-mail address: jqshen@coer.zju.edu.cn}}
\address{$^{1}$  Centre for Optical
and Electromagnetic Research, State Key Laboratory of Modern
Optical Instrumentation, Zhejiang University,
Hangzhou SpringJade 310027, P. R. China\\
$^{2}$Zhejiang Institute of Modern Physics and Department of
Physics, Zhejiang University, Hangzhou 310027, P. R. China}
\date{\today }
\title{A Supplement: on the Quantum-vacuum Geometric Phases\footnote{The present paper is a supplement to the two papers: J.Q. Shen and H.Y. Zhu, Ann.Phys.(Leipzig) {\bf
12} (2003) p.131 [3] and arXiv: quant-ph/0309007 [6](entitled: An
experimental realization of quantum-vacuum geometric phases by
using the gyrotropic-medium optical fiber). It will be submitted
nowhere else for the publication, just uploaded at the e-print
archives.}} \maketitle

\begin{abstract}
Three related topics on the quantum-vacuum geometric phases inside
a noncoplanarly curved optical fiber is presented: (i) a brief
review: the investigation of vacuum effect and its experimental
realization; (ii) our sequence of ideas of geometric phases of
photons in the fiber; (iii) three derivations of effective
Hamiltonian that describes the wave propagation of photon field in
a curved fiber.
\end{abstract}
\section{A brief review: the investigation of vacuum effect\\ and its experimental realization}
Geometric phases has received special attention of many
researchers since Berry showed that there exists a topological
phase ({\it i.e.}, the adiabatic geometric phase) in
quantum-mechanical wavefunction in adiabatic quantum process where
the cyclic evolution of wavefunction yields the original state
plus a phase shift, which is a sum of a dynamical phase and a
geometric phase shift\cite {Berry}. In the published
papers\cite{Gao,Shen1} Gao and we considered a second-quantized
spin model which can characterize the wave propagation of photon
field in a noncoplanarly curved optical fiber\cite{Chiao,Tomita}.
In this work\cite{Gao,Shen1}, we dealt only with the case where
the photon spin operator is taken the normal-order form, which
does not involve the vacuum zero-point electromagnetic fluctuation
fields. So, the previous mathematical treatment\cite{Gao,Shen1}
cannot predict the existence of the geometric phases resulting
from the vacuum photon fluctuation in the curved fiber. Here we
refer to such vacuum-induced topological phases as the {\it
quantum-vacuum geometric phases}. In order to treat the so-called
{\it quantum-vacuum geometric phases}, we should study the
non-normal-order spin operators of photon field, where the
zero-point electromagnetic fields is involved in the effective
Hamiltonian (and hence in the time-evolution equation, {\it i.e.},
the time-dependent Schr\"{o}dinger equation). For the detailed
discussions, readers may be referred to the paper\cite{Shen2},
where we discussed the contribution of vacuum quantum fluctuation
to geometric phases of photons moving inside a sufficiently
perfect optical fiber and suggested an experimental scheme to
detect them by using the noncoplanar fiber made of certain
gyrotropic materials with suitable permittivity and permeability
tensors\cite{Shen2}.

Nearly four years has passed since we began to consider the
quantum-vacuum geometric phases. The reason that drew our
attention to such vacuum effects was as follows: in the quantum
field theory the infinite zero-point vacuum energy is often
cancelled (deleted) by using the normal-product procedure and a
new vacuum background is thus re-defined\cite{Bjorken2}. It is
believed that this formalism is valid for the {\it
time-independent} quantum systems since in these cases the
divergent background energies may have no observable effects
(except for the Casimir's effect\cite{Casimir}, vacuum
polarization leading to Lamb's shift\cite{Bethe}, atomic
spontaneous radiation due to the interaction of the excited atom
with the zero-point electromagnetic field, as well as the
anomalous magnetic moment of electron) and hence do not influence
the observed physical results of the interacting quantum fields.
However, if the normal product technique is applied to the {\it
time-dependent} systems of quantum field theory (such as field
theory in the expanding universe and the time-dependent
gravitational backgrounds\cite{Fulling,Ford,Fu}), and thus the
vacuum background is so re-defined by removing {\it different}
zero-point energies at {\it different} time, then some observable
vacuum effects ({\it e.g.}, the vacuum contribution to Berry's
phase that arises in time-dependent quantum systems) may be
deleted theoretically and therefore the validity of this formalism
may deserve incredulity\footnote{If in experiments we find no
information on the existence of this geometric phases at
quantum-vacuum level ({\it i.e.}, this vacuum effect does not
exist), then it is believed that the normal-product procedure in
second quantization is still valid for {\it time-dependent}
quantum systems. But, once if the quantum-vacuum geometric phases
is truly present experimentally, then we might argue that the
normal-product procedure in second quantization may be no longer
valid for {\it time-dependent} quantum systems, namely, these
vacuum effects associated with the time-dependent evolution
process in the time-dependent quantum field theory should not be
removed by the normal-product procedure ({\it i.e.}, it is not
appropriate to re-define a vacuum background by removing {\it
different} zero-point energies at {\it different} time).}.

Unfortunately, the left-handed polarized light due to vacuum
fluctuation always inescapably coexists with the zero-point
right-handed polarized light and the total of quantum-vacuum
geometric phases is therefore vanishing\cite{Shen2}, which may
follow from the following expressions for the cyclic adiabatic
geometric phases\cite{Shen2}
\begin{equation}
 \phi_{L}=-\left(n_{L}+\frac{1}{2}\right)\cdot2\pi\left(1-\cos\lambda\right),   \quad
 \phi_{R}=\left(n_{R}+\frac{1}{2}\right)\cdot2\pi\left(1-\cos\lambda\right),              \label{1}
\end{equation}
where $n_{L}$ and $n_{R}$ denote the occupation numbers of left-
and right- handed (LRH) circularly polarized photons,
respectively. Note that the geometric phases at quantum-vacuum
level of left-handed polarized light,
\begin{equation}
\phi^{({\rm
vacuum})}_{L}=-\frac{1}{2}\cdot2\pi\left(1-\cos\lambda\right),
\end{equation}
 acquires
a minus sign, which may be cancelled (counteracted) by $\phi^{(\rm
vacuum)}_{R}$, {\it i.e.},
\begin{equation}
\phi^{(\rm
vacuum)}_{R}=+\frac{1}{2}\cdot2\pi\left(1-\cos\lambda\right)
\end{equation}
 that possesses a plus
sign\cite{Shen2}.

Since the total vacuum geometric phases ($\phi^{({\rm
vacuum})}+\phi^{(\rm vacuum)}_{R}$) is vanishing and therefore
trivial, it prevents physicists from investigating experimentally
this nontrivial vacuum effect ({\it i.e.}, vacuum contribution to
the geometric phases). This, therefore, means that our above
theoretical remarks as to whether the normal-product procedure is
valid or not for the {\it time-dependent} quantum field theory
(TDQFT) cannot be easily examined experimentally. It depressed me
to be confronted with such difficulties.

Moreover, as was stated more recently by Fuentes-Guridi {\it et
al.}, in a strict sense, the Berry phase has been studied only in
a semiclassical context until now\cite{Fuentes}. Thus the effects
of the vacuum field on the geometric evolution are still
unknown\cite{Fuentes}. So, we think that in addition to its
physical significance in investigating the normal-product
procedure in {time-dependent} field theory, the quantum-vacuum
geometric phases itself is also physically interesting. These
factors always encouraged me to seek a way to test this vacuum
effect.

During the last four years, I tried my best but often
unfortunately failed to suggest an excellent idea of experimental
realization of this quantum-vacuum geometric phases of photons in
the fiber. We conclude that it seems not quite satisfactory to
test the quantum-vacuum geometric phases by using the optical
fiber that is made of isotropic media, inhomogeneous media ({\it
e.g.}, photonic crystals\footnote{Photonic crystals are artificial
materials patterned with a periodicity in dielectric constant,
which can create a range of forbidden frequencies called a
photonic band gap. Such dielectric structure of crystals offers
the possibility of molding the flow of light (including the
zero-point electromagnetic fields of vacuum). It is believed that,
in the similar fashion, this effect ({\it i.e.}, modifying the
mode structures of vacuum electromagnetic fields) may also take
place in gyrotropic media. The suppression of spontaneous emission
in certain gyrotropic media can be regarded as an illustrative
example.}), left-handed media (a kind of artificial composite
metamaterial with the negative permittivity and permeability and
the consequent negative refractive index\cite{Veselago,Pendry3}),
uniaxial (biaxial) crystals or chiral materials. Is it truly
extremely difficult to realize such a goal? It is found finally
that perhaps in the fiber composed of some anisotropic media such
as gyrotropic materials (gyroelectric or gyromagnetic media) the
quantum-vacuum geometric phases may be achieved test
experimentally.

Gyrotropic media is such electromagnetic materials where both the
electric permittivity and the magnetic permeability are tensors,
which can be respectively written as\cite{Veselago}
\begin{equation}
(\epsilon)_{ik}=\left(\begin{array}{cccc}
\epsilon_{1}  & i\epsilon_{2} & 0 \\
-i\epsilon_{2} &   \epsilon_{1} & 0  \\
 0 &  0 &  \epsilon_{3}
 \end{array}
 \right),                 \qquad          (\mu)_{ik}=\left(\begin{array}{cccc}
\mu_{1}  & i\mu_{2} & 0 \\
-i\mu_{2} &   \mu_{1} & 0  \\
 0 &  0 &  \mu_{3}
 \end{array}
 \right) .              \label{2}
\end{equation}
Let us assume that the electromagnetic wave is propagating along
the third component of the Cartesian coordinate system. By using
the Maxwellian equations, it is verified that the optical
refractive indices squared of gyrotropic media corresponding to
the two directions of polarization vectors (left- and right-
handed polarized) are of the form
$n^{2}_{\pm}=(\epsilon_{1}\pm\epsilon_{2})(\mu_{1}\pm\mu_{2})$.
According to this expression for $n^{2}_{\pm}$, it is possible for
us to choose: $n^{2}_{+}>0$ and $n^{2}_{-}<0$\cite{Veselago}, and
then in these gyrotropic media only one of the LRH polarized
lights can be propagated without being absorbed by media. This
result holds also for the zero-point electromagnetic
fields\cite{Yablonovitch,Hulet,Jhe,Meshede} ({\it e.g.}, the
inhibition and enhancement of spontaneous emission in photonic
crystals and cavity resonator, where the vacuum mode density is
modified under certain conditions). If, for example, in some
certain gyrotropic media one of the LRH polarized lights, say, the
left-handed polarized light (including its vacuum fluctuation
field), dissipates due to the medium absorption and only the
right-handed light is allowed to be propagated (in the meanwhile
the vacuum mode structure in these anisotropic media also alters
correspondingly), then only the quantum-vacuum geometric phase of
right-handed polarized light is retained and can be easily tested
in the fiber fabricated from these gyrotropic media\cite{Shen2}.

Since it is known that people can manipulate vacuum so as to alter
the zero-point mode structures of vacuum, which has been
illustrated in photonic crystals\cite{Yablonovitch} and Casimir's
effect (additionally, the space between two parallel mirrors,
cavity resonator in cavity QED\cite{Hulet,Jhe,Meshede}, {\it
etc.})\footnote{Spontaneous emission from an excited electronic
state reflects the properties of the surrounding vacuum-field
fluctuations. By placing the radiator near a metallic surface or
in a cavity, one can modify the spectral density of these
fluctuations and either enhance or inhibit spontaneous emission
[18]. For instance, spontaneous radiation by an atom in a Rydberg
state has been inhibited by use of parallel conducting planes to
eliminate the vacuum modes at the transition frequency [17].}, we
will put forward another scheme to extract the nontrivial
quantum-vacuum geometric phases in the noncoplanar fiber by using
Casimir's effect. It is well known that in a finitely large space
({\it e.g.}, the space between two parallel metallic plates, which
is the main equipment of Casimir's effect experiment), the
vacuum-fluctuation electromagnetic field alters its mode
structures\footnote{Because its ``wavelength'' is much larger than
the length scale of the surrounding space, it cannot form a
stationary wave.}, namely, the zero-point field with wave vector
$k$ less than $\sim\left(\frac{\pi}{a}\right)$ is expelled from
this space with a finite scale length $a$. Likewise, in some
certain gyrotropic media, if the electromagnetic parameters
$\epsilon_{1}$, $\epsilon_{2}$ and $\mu_{1}$, $\mu_{2}$ in
expressions (\ref{2}) for electric permittivity and magnetic
permeability tensors can be chosen to be
$\epsilon_{1}\simeq\epsilon_{2}$ and $\mu_{1}\simeq\mu_{2}$, then
the optical refractive index $n_{-}$ of left-handed polarized
light is very small (or approaching zero) and hence the wave
vector $k_{-}\simeq 0$ , since in these media $k_{-}$ is
proportional to $n_{-}$. Thus the left-handed polarized zero-point
field inside these gyrotropic media is absent if, for example, the
media are placed in a finitely large space, and consequently the
only retained quantum-vacuum geometric phase is that of
right-handed circularly polarized light. In order to perform the
detection of this geometric phase, we should make use of the
optical fiber that is fabricated from the above gyrotropic media,
and the devices used in Tomita-Chiao fiber
experiments\cite{Chiao,Tomita} should be placed in a sealed
metallic chamber or cell, where the zero-point circularly
polarized field with lower wave vector does not exist. Hence the
measurement of quantum-vacuum geometric phases of photons may be
achievable by means of this scheme.

Although the infinite vacuum energy in conventional {\it
time-independent} quantum field theories is harmless and easily
removed theoretically by normal-order procedure, here for a {\it
time-dependent} quantized-field system, we think that the
existence of quantum-vacuum geometric phases indicates that
zero-point fields of vacuum will also participate in the time
evolution process and perhaps can no longer be regarded merely as
an inactive onlooker in {\it time-dependent} quantum field
theories such as field theory in curved space-time, {\it e.g.}, in
time-dependent gravitational backgrounds and expanding
universe\cite{Fulling,Ford}. In order to investigate this
fundamental problem of quantum field theory, we hope the vacuum
effect presented here would be tested experimentally in the near
future. We also hope to see anyone else putting forward some new
more clever suggestions of detecting this interesting geometric
phases at quantum-vacuum level.
\\ \\
{\bf Note added}: Just after I finished the paper\cite{Shen2},
S.L. Zhu of Hongkong University drew my attention to
Fuentes-Guridi {\it et al.}'s recently published
work\cite{Fuentes}, where they considered the interaction between
a spin-$1/2$ charged particle (similar to the case of two-level
atom) and the external quantized electromagnetic field, and
studied a so-called vacuum-induced Berry's phase which they
regarded as the contribution of vacuum-field fluctuation. I think
that Fuentes-Guridi {\it et al.}'s result can also be treated in
more detail by the method presented in our paper\cite{Shen1},
where we obtained the exact solutions of the time-dependent
supersymmetric two-level multiphoton Jaynes-Cummings model.

\section{Our sequence of ideas of investigation of geometric phases in the fiber}
During the past four years, we studied the noncyclic nonadiabatic
time evolution of photon wavefunction in the noncoplanar fiber by
using the Lewis-Riesenfeld invariant theory, second-quantized spin
model and non-normal product procedure\cite{Gao,Shen1,Shen2}. The
outline of our sequence of ideas is given as follows:
\\ \\
 \setlength{\unitlength}{1cm}
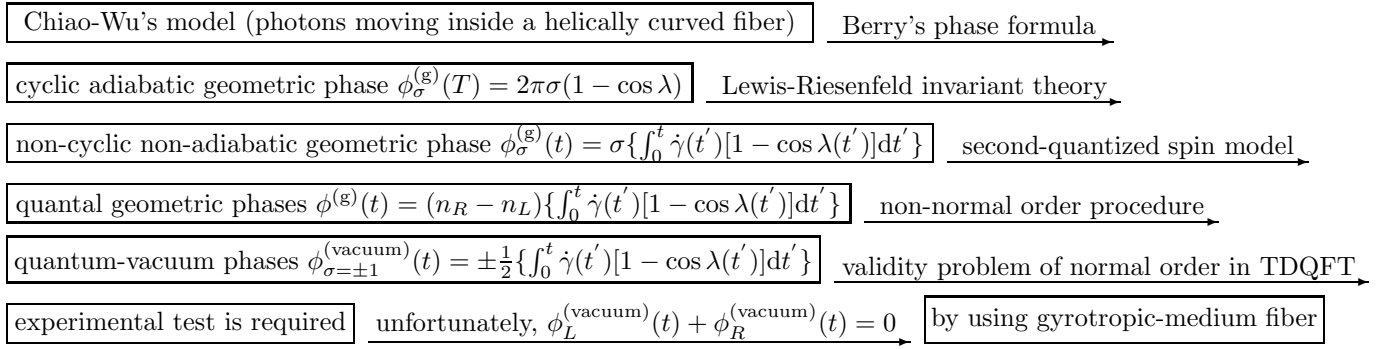
\begin{picture}(30,5)
\put(0,4){\framebox(10.7,0.5){Chiao-Wu's model (photons moving
inside a helically curved fiber)}} \put(10.9,4){\vector(1,0){3.8}}
\put(11.1,4.1){Berry's phase formula}
\put(0,3.2){\framebox(9.1,0.5){cyclic adiabatic geometric phase
$\phi _{\sigma }^{\rm (g)}(T)=2\pi\sigma(1-\cos \lambda )$}}
\put(9.3,3.2){\vector(1,0){5.5}} \put(9.5,3.3){Lewis-Riesenfeld
invariant theory} \put(0,2.4){\framebox(12.3,0.5){non-cyclic
non-adiabatic geometric phase $\phi _{\sigma }^{\rm
(g)}(t)=\sigma\{{\int_{0}^{t}\dot{\gamma}(t^{^{\prime }})[1-\cos
\lambda (t^{^{\prime }})]{\rm d}t^{^{\prime }}}\}$}}
\put(12.5,2.4){\vector(1,0){4.8}} \put(12.7,2.5){second-quantized
spin model}

\put(0,1.6){\framebox(11.2,0.5){quantal geometric phases
$\phi^{\rm
(g)}(t)=(n_{R}-n_{L})\{{\int_{0}^{t}\dot{\gamma}(t^{^{\prime
}})[1-\cos \lambda (t^{^{\prime }})]{\rm d}t^{^{\prime }}}\}$}}
\put(11.4,1.6){\vector(1,0){4.7}} \put(11.6,1.7){non-normal order
procedure}

\put(0,0.8){\framebox(10.8,0.6){quantum-vacuum phases
$\phi_{\sigma=\pm 1}^{\rm
(vacuum)}(t)=\pm\frac{1}{2}\{{\int_{0}^{t}\dot{\gamma}(t^{^{\prime
}})[1-\cos \lambda (t^{^{\prime }})]{\rm d}t^{^{\prime }}}\}$}}
\put(11,0.8){\vector(1,0){7.1}} \put(11.1,0.9){validity problem of
normal order in TDQFT}

\put(0,0){\framebox(4.6,0.5){experimental test is required}}
\put(4.8,0){\vector(1,0){7.2}} \put(4.9,0.15){unfortunately,
$\phi_{L}^{\rm (vacuum)}(t)+\phi_{R}^{\rm (vacuum)}(t)=0$
}\put(12.2,0){\framebox(5.3,0.6){by using gyrotropic-medium
fiber}}
\end{picture}

\section{Three derivations of effective Hamiltonian of photons inside a
noncoplanarly curved fiber}
 The effective Hamiltonian describing
the light wave propagation in a curved optical fiber is helpful in
considering the nonadiabatic noncyclic time evolution process of
photon wavefunction in the fiber. We have three methods to derive
this effective Hamiltonian\cite{Gao,Shen1}
\begin{equation}
H_{\rm eff}(t)=\frac{%
{\bf{k}}(t)\times \dot{\bf{k}}(t)}{k^{2}}\cdot {\bf{S}},
\label{eq220}
\end{equation}
where dot denotes the derivative of wave vector ${\bf k}$ with
respect to time and ${\bf{S}}$ stands for the photon spin
operators.
\\

{\bf Method i} \quad   By using the infinitesimal rotation
operator of wavefunction

The photon wavefunction $|\sigma, {\bf k}(t)\rangle$ varies as it
rotates by an infinitesimal angle, say $\vec{\vartheta}$, namely,
it obeys the following transformation rule
\begin{equation}
|\sigma, {\bf
k}(t+\Delta{t})\rangle=\exp\left[-i\vec{\vartheta}\cdot{\bf
J}\right]|\sigma, {\bf k}(t)\rangle,     \label{eqA1}
\end{equation}
where $\exp\left[-i\vec{\vartheta}\cdot{\bf J}\right]\simeq
1-i\vec{\vartheta}\cdot{\bf J}$ with ${\bf J}$ being the total
angular momentum operator of photon and ${\bf k}(t+\Delta{t})={\bf
k}(t)+\Delta {\bf k}(t)$ with $\Delta {\bf
k}(t)=\dot{\bf{k}}{\Delta t}$. Here $|\vec{\vartheta}|$ is the
angle displacement vector between ${\bf k}(t)$ and ${\bf
k}(t+\Delta{t})$, and the direction of $\vec{\vartheta}$ is
parallel to that of ${\bf k}(t)\times{\bf k}(t+\Delta{t})$. One
can therefore arrive at
\begin{equation}
\vec{\vartheta}=\frac{{\bf k}(t)\times{\bf
k}(t+\Delta{t})}{k^2}=\frac{{\bf
k}(t)\times\dot{\bf{k}}}{k^2}{\Delta t}.
 \label{eqApp2}
\end{equation}
Thus it follows from Eq.(\ref{eqA1}) and (\ref{eqApp2}) that
\begin{equation}
i\frac{\partial \left| \sigma ,{\bf{k}}(t)\right\rangle }{\partial
t}=\frac{{\bf{k}}(t)\times \dot{\bf{k}}(t)}{k^{2}}\cdot
{\bf{J}}\left| \sigma ,{\bf{k}}(t)\right\rangle \label{eqA3}
\end{equation}
by calculating the time derivative of $|\sigma, {\bf k}(t+\Delta
{t})\rangle$. The total angular momentum is
${\bf{J}}={\bf{L}}+{\bf{S}}$, where the orbital angular momentum
${\bf{L}}$ is orthogonal to the linear momentum ${\bf{k}}$ for the
photon. So, $\frac{{\bf{k}}(t)\times \dot{\bf{k}}(t)}{k^{2}}\cdot
{\bf{L}}=0$ and the only retained term in $\frac{{\bf{k}}(t)\times
\dot{\bf{k}}(t)}{k^{2}}\cdot {\bf{J}}$ is $\frac{{\bf{k}}(t)\times
\dot{\bf{k}}(t)}{k^{2}}\cdot {\bf{S}}$. This, therefore, means
that if we think of Eq.(\ref{eqA3}) as the time-dependent
Schr\"{o}dinger equation governing the propagation of photons in
the noncoplanar fiber, then we can obtain the effective
Hamiltonian (\ref{eq220}).
\\

{\bf Method ii} \quad  By using the equation of motion of a photon
in a ``gravitomagnetic'' field

If the momentum squared ${\bf k}^2$ of a photon moving in a
noncoplanarly curved and sufficiently perfect optical fiber is
conserved, then we can derive the following identity
\begin{equation}
\dot{\bf{k}}+{\bf{k}}\times \left(\frac{{\bf{k}}\times
\dot{\bf{k}}}{k^{2}}\right)=0,            \label{A4}
\end{equation}
which can be regarded as the equation of motion of a photon in the
noncoplanarly curved fiber. Since Eq.(\ref{A4}) is exactly
analogous to the equations of motion of a charged particle moving
in a magnetic field or a spinning particle moving in a rotating
frame of reference, $-\frac{{\bf{k}}\times \dot{\bf{k}}}{k^{2}}$
can be considered a ``magnetic field'' or ``gravitomagnetic
field'' (thus $-{\bf{k}}\times (\frac{{\bf{k}}\times
\dot{\bf{k}}}{k^{2}})$ can be thought of as a ``Lorentz magnetic
force'' or ``Coriolis force''). Similar to the Mashhoon {\it et
al.}'s work ({\it i.e.}, the derivation of the interaction
Hamiltonian of gravitomagnetic dipole moment in a gravitomagnetic
field)\cite{Shen3,Mashhoon}, one can also readily write the
Hamiltonian describing the coupling of the photon
``gravitomagnetic moment'' ({\it i.e.}, photon spin ${\bf S
}$)\cite{Shen2} to the ``gravitomagnetic field'' as follows
\begin{equation}
H=\frac{{\bf{k}}\times \dot{\bf{k}}}{k^{2}}\cdot {\bf{S}},
 \label{A5}
\end{equation}
which is just the expression (\ref{eq220}).
\\

{\bf Method iii} \quad   By using the Liouville-Von Neumann
equation

If a photon is moving inside a noncoplanarly curved optical fiber
that is wound smoothly on a large enough diameter\cite{Tomita},
then its helicity reversal does not easily take place\cite{Guo}
and the photon helicity $I(t)=\frac{{\bf k}(t)}{k}\cdot{\bf S}$ is
therefore conserved\cite{Chiao} and can thus be considered a
Lewis-Riesenfeld invariant $I(t)$\cite{Lewis}, which agrees with
the Liouville-Von Neumann equation
\begin{equation}
\frac{\partial I(t)}{\partial t}+\frac{1}{i}[I(t), H(t)]=0.
\label{eq5}
\end{equation}
With the help of the spin operator commuting relations ${\bf
S}\times {\bf S}=i{\bf S}$, one can solve the Liouville-Von
Neumann equation (\ref{eq5}), namely, if the effective
Hamiltonian\footnote{Since in the Liouville-Von Neumann equation,
$I(t)$ is $\frac{{\bf k}(t)}{k}\cdot{\bf S}$, it is certain that
$H(t)$ should also be the linear combination of photon spin
operator $S_{1}, S_{2}, S_{3}$.} is written as $H(t)={\bf
h}(t)\cdot{\bf S}$, then according to the Liouville-Von Neumann
equation, one can arrive at
$\left[I(t),H(t)\right]=\left[\frac{{\bf k}(t)}{k}\times{\bf
h}(t)\right]\cdot i{\bf S}$, and readily obtain the expression for
the effective Hamiltonian (\ref{eq220}) of photons in the curved
optical fiber, {\it i.e.}, the coefficients of the effective
Hamiltonian is ${\bf h}(t)=\frac{{\bf{k}}(t)\times
\dot{\bf{k}}(t)}{k^{2}}$.

It is apparently seen that substitution of $I(t)=\frac{{\bf
k}(t)}{k}\cdot{\bf S}$ and $H(t)=\frac{{\bf{k}}(t)\times
\dot{\bf{k}}(t)}{k^{2}}\cdot {\bf{S}}$ into the Liouville-Von
Neumann equation (\ref{eq5}) yields the equation of motion of a
photon in a ``gravitomagnetic'' field, {\it i.e.}, Eq.(\ref{A4}).
\\ \\
\textbf{Acknowledgements}  This project is supported by the
National Natural Science Foundation of China under the project No.
$90101024$. I benefited much from useful advice of X.C. Gao when I
tried my best to suggest an experimental scheme to extract the
nontrivial quantum-vacuum geometric phases in the noncoplanar
fiber. My special thanks are due to him also.

\end{document}